\documentclass[pra,twocolumn,showpacs]{revtex4}
\usepackage{amsmath,amssymb,mathrsfs}
\usepackage{graphicx}
\usepackage{amsmath}
\usepackage{epsfig}
\usepackage[usenames]{color}

\newcommand{\beq}{\begin{equation}}
\newcommand{\eeq}{\end{equation}}
\newcommand{\beqa}{\begin{eqnarray}}
\newcommand{\eeqa}{\end{eqnarray}}
\newcommand{\ba}{\begin{array}}
\newcommand{\ea}{\end{array}}

\begin{document}

\title{Collective modes in the anisotropic unitary Fermi gas \\ 
and the inclusion of a backflow term}

\author{L. Salasnich, P. Comaron, M. Zambon, F. Toigo}
\affiliation{Dipartimento di Fisica e Astronomia ``Galileo Galilei'' and 
CNISM, Universit\`a di Padova, Via Marzolo 8, 35131 Padova, Italy}

\date{\today}

\begin{abstract}
We study the collective modes of the confined unitary Fermi gas 
under anisotropic harmonic confinement as a function of 
the number of atoms. We use the equations of extended 
superfluid hydrodynamics, which take into account 
a dispersive von Weizs\"acker-like term in the equaton of state. 
Finally, we discuss the inclusion of a backflow term 
in the extended superfluid Lagrangian and the effects of this 
anomalous term on sound waves and Beliaev damping of phonons.  
\end{abstract}


\maketitle

\section{Introduction}

In this paper we calculate the 
collective monopole and quadrupole modes of 
the unitary Fermi gas (characterized by an 
infinite s-wave scattering length) under axially-symmetric anisotropic 
harmonic confinement by using the extended Lagrangian density 
of superfluids which we proposed a few years ago \cite{flavio},  
to study the unitary Fermi gas 
\cite{flavio,flavio2,flavio3,flavio4,flavio5,flavio6,flavio7,flavio8}. 
The internal energy density of our extended Lagrangian density 
contains a term proportional to the kinetic 
energy of a uniform non interacting gas of fermions, plus 
a gradient correction of the von-Weizsacker 
form $\lambda \hbar^2/(8m) (\nabla n/n)^2$ \cite{von}. 
The inclusion of a gradient term has been adopted for studying 
the quantum hydrodynamics of electrons by
March and Tosi \cite{tosi}, and by Zaremba and Tso \cite{tso}.
In the context of the BCS-BEC crossover, the gradient term 
is quite standard \cite{nick,kim,v2,v3,v4,v5,v6,v7,v8}. 
In the last part of this paper we consider the inclusion of 
backflow terms \cite{less,treiner} 
in the extended superfluid Lagrangian. By using our equations 
of extended superfluid hydrodynamics with backflow 
we calculate sound waves, static response function and 
structure factor of a generic uniform superfluid and 
also the effect of the backflow on Beliaev damping 
of phonons \cite{beliaev}. 

\section{Extended superfluid Lagrangian and hydrodynamic equations}

The extended Lagrangian density of dilute and ultracold 
superfluids is given 
by \cite{flavio,flavio2,flavio3,flavio4,flavio5,flavio6,flavio7,flavio8} 
\beq 
\label{lagrangian}
\mathscr{L} = \mathscr{L}_0 + \mathscr{L}_W
\eeq
where 
\beq
\mathscr{L}_0 = - \hbar \, {\dot \theta} \, n - 
{\hbar^2\over 2m} ({\boldsymbol \nabla} \theta)^2 \, n 
- U({\bf r}) \, n - {\cal E}_0(n)
\label{l0}
\eeq
is the familiar Popov's Lagrangian density \cite{popov} 
of superfluid hydrodynamics,  
with $n({\bf r},t)$ the local density and 
$\theta({\bf r},t)$ half of the phase of the 
condensate order parameter of Cooper pairs for  
superfluid fermions (or the phase of the condensate order parameter 
for superfluid bosons). Here $U({\bf r})$ is the 
external potential acting on particles and $ {\cal E}_0(n)$ is the 
bulk internal energy of the system. The generalization of the 
superfluid hydrodynamics is due to the the Lagrangian density 
\beq
\mathscr{L}_W = 
- \lambda {\hbar^2\over 8m} {({\boldsymbol\nabla} n)^2\over n} \; , 
\label{lw}
\eeq
which takes into account density variations. 
Thus, the local internal energy 
depends not only 
on the local density  $n({\bf r},t)$ but also on its space gradient,  
namely  
\beq 
{\cal E}(n,{\boldsymbol\nabla} n) = {\cal E}_0(n) + 
\lambda {\hbar^2\over 8m} {({\boldsymbol\nabla} n)^2\over n} \; , 
\eeq 
where, as previously mentioned, ${\cal E}_0(n)$ is the internal 
energy of a uniform unitary Fermi gas with density $n$.
The parameter $\lambda$ giving the gradient correction 
must be determined from microscopic calculations or 
from comparison with experimental data. 

By using the Lagrangian density (\ref{lagrangian})  
the Euler-Lagrange equation for $ \theta$ gives 
\beq 
{\partial n \over \partial t} + {\hbar\over m} {\boldsymbol\nabla} \cdot 
\left( n \; {\boldsymbol\nabla} \theta \right) = 0 \; , 
\label{el1}
\eeq
while the Euler-Lagrange equation for $n$ leads to
\beq 
\hbar \, {\dot \theta} + {\hbar^2\over 2m}({\boldsymbol\nabla}\theta)^2 
+ U({\bf r}) + X(n,{\boldsymbol \nabla}n) = 0 \; ,  
\label{careful}
\eeq
where 
\beq 
X(n,{\boldsymbol \nabla}n) = 
{\partial {\cal E} \over \partial n} -  
{\boldsymbol\nabla}\cdot
{\partial {\cal E} \over \partial ({\boldsymbol\nabla} n)} \; . 
\label{x-def}
\eeq
which describes how the internal energy varies as the local density 
and its gradient vary, may be considered a local chemical potential.
The local velocity field ${\bf v}({\bf r},t)$ of the superfluid is 
related to $\theta({\bf r},t)$ by 
\beq 
{\bf v}({\bf r},t) = {\hbar \over m} 
{\boldsymbol \nabla} \theta({\bf r},t) \; . 
\label{velocity}
\eeq
This definition ensures that the velocity is irrotational, i.e. 
${\boldsymbol\nabla} \wedge {\bf v} = {\bf 0}$. 
By using the definition (\ref{velocity}) in both Eqs. (\ref{el1}) 
and (\ref{careful}) 
and applying the gradient operator $\boldsymbol\nabla$ 
to Eq. (\ref{careful}) one finds the extended hydrodynamic 
equations of superfluids
\beqa 
{\partial n \over \partial t} + {\boldsymbol\nabla} \cdot 
\left( n \; {\bf v} \right) = 0 \; . 
\label{continuo}
\\
m {\partial {\bf v} \over \partial t} + 
{\boldsymbol\nabla} \left[ {1\over 2} m {\bf v}^2 + U({\bf r}) + 
X(n,{\boldsymbol \nabla} n) \right] = {\bf 0} \; . 
\label{motion}
\eeqa
We stress that in the presence of an external confinement $U({\bf r})$ 
the chemical potential $\mu$ of the system does not coincide 
with the local chemical potential $X(n,{\boldsymbol \nabla} n)$.
In the presence of an external potential the relation between 
the equilibrium (ground state) density $n_0 ({\bf r})$ 
and the chemical potential $\mu$ can be obtained from Eq. (\ref{careful}) 
by setting $\theta({\bf r},t)=-\mu t/\hbar$ 
and ${\bf v}({\bf r},t)={\bf 0}$, so that 
\beq  
U({\bf r}) + X(n_0,{\boldsymbol \nabla}n_0) = \mu \; . 
\label{ground-state}
\eeq

\section{Collective modes of the anisotropic unitary Fermi gas}

In the case of the unitary Fermi gas 
the bulk internal energy can be written as 
\beq 
{\cal E}_0(n) = \xi \, {3\over 5} {\hbar^2\over 2m} 
(3\pi^2)^{2/3} \, n^{5/3} \; ,  
\label{e00}
\eeq
where $\xi\simeq 0.4$ is a universal parameter 
\cite{flavio,flavio2,v4,son,valle}. 
and various approaches 
\cite{flavio,flavio2,v4,v7,son,valle} 
suggest that $\lambda \simeq 0.25$. The local chemical 
potential is then:
\beq 
X(n,{\boldsymbol\nabla} n) =  {\hbar^2\over 2m} 
(3\pi^2)^{2/3} \xi \, n^{2/3} -  
\lambda {\hbar^2\over 2m}
{\nabla^2\sqrt{n}\over \sqrt{n}} \; .  
\label{invert}
\eeq
with the above mentioned values of $ \xi$ and $\lambda$.

In this section we consider the unitary Fermi gas 
under the anisotropic axially-symmetric harmonic confinement
\beq
U({\bf r})= \frac{m}{2} \omega^2_{\rho} (x^2+y^2)
+ \frac{m}{2} \omega_z^2 z^2 \; ,  
\label{oscillator}
\eeq
where $\omega_{\rho}$ is the cylindric radial frequency while 
$\omega_z$ is the axial frequency. 
In this case, Eq. (\ref{ground-state}) for the ground-state 
density profile $n_0({\bf r})$ becomes 
\beqa  
\frac{m}{2} \omega^2_{\rho} (x^2+y^2)
&+& \frac{m}{2} \omega_z^2 z^2 + 
{\hbar^2\over 2m} 
(3\pi^2)^{2/3} \xi \, n_0(x,y,z)^{2/3} 
\nonumber
\\
&-&  
\lambda {\hbar^2\over 2m} {\nabla^2 \sqrt{n_0(x,y,z)}
\over \sqrt{n_0(x,y,z)}} 
= \mu \; . 
\eeqa
We have solved numerically this 3D partial differential equation, 
by using a finite-difference predictor-corrector 
Crank-Nicholson method \cite{sala-numerics} with imaginary 
time after chosing $\xi=0.42$ and $\lambda=0.25$.
In the case of isotropic trap ($\omega_{\rho}/\omega_z=1$) 
the fermionic cloud is spherically symmetric and consequently 
axial and radial density profiles coincide. Instead, as expected, 
by increasing the trap anisotropy also the fermionic cloud 
becomes more anisotropic.

We are interested in calculating the frequencies of 
low-lying collective oscillations of the anisotropic unitary Fermi gas. 
Exact scaling solutions for the unitary Fermi gas have been considered 
by Castin \cite{castin} and also by Hou, Pitaevskii, and Stringari 
\cite{forzati}. Unfortunately, in the presence of anisotropic 
trapping potential and including the gradient 
term in the hydrodynamic equations, these scaling solutions 
are no more exact. 

\begin{figure}[tbp]
\begin{center}
{\includegraphics[width=9.cm,clip]{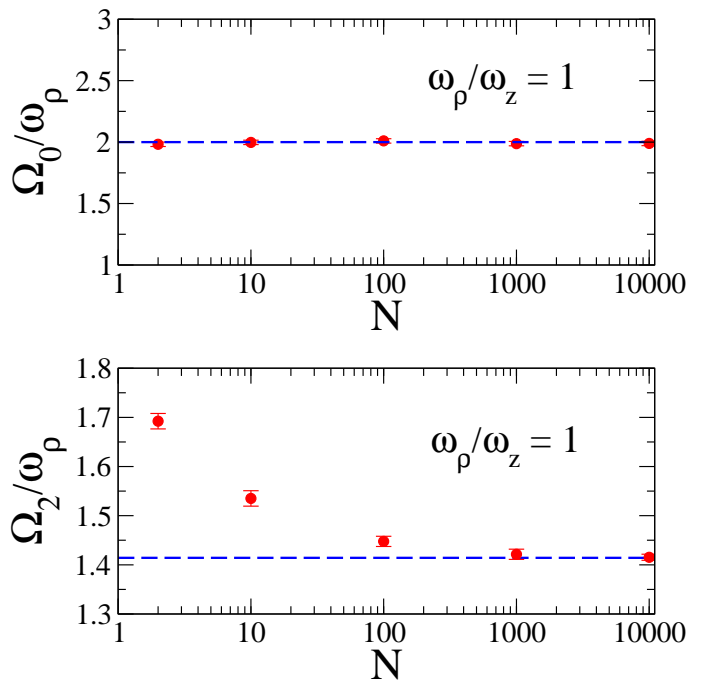}}
\end{center}
\caption{(Color online). Unitary Fermi gas under isotropic 
($\omega_{\rho}=\omega_z$) harmonic confinement. 
In the two panels there are the monopole frequency 
$\Omega_{0}$ (upper panel) and the quadrupole frequency 
$\Omega_{2}$ (lower panel) as a function of the number $N$ of atoms. 
Filled circles with error bars: numerical results 
obtained solving Eqs. (\ref{continuo}) and (\ref{motion}) 
with Eq. (\ref{invert}) and $\lambda=0.25$. 
Dashed lines: analytical results, i.e. exact Eq. (\ref{exact-castin}) 
and Thomas-Fermi Eq. (\ref{tf-stringa}). 
Universal parameter of the unitary Fermi gas: $\xi=0.42$.}
\label{fig1}
\end{figure}

\begin{figure}[tbp]
\begin{center}
{\includegraphics[width=9.cm,clip]{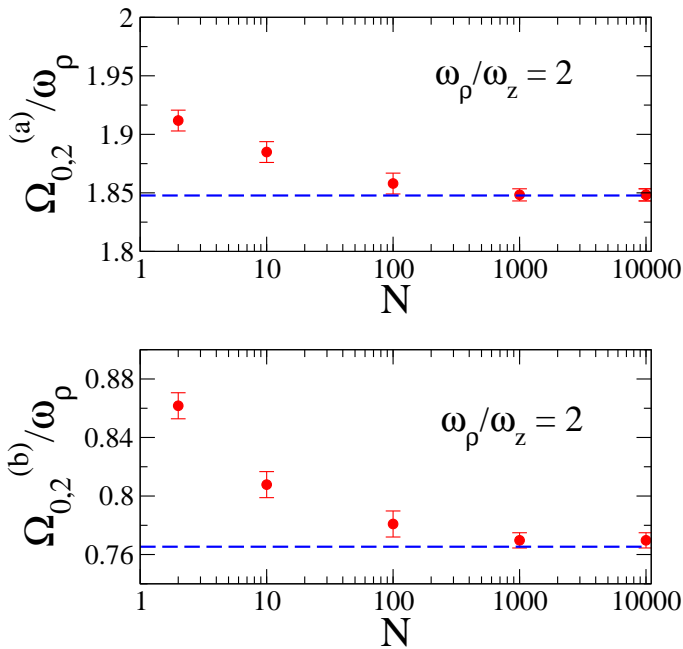}}
\end{center}
\caption{(Color online). Unitary Fermi gas under anisotropic 
but axially-symmetric ($\omega_{\rho}=2\omega_z$) harmonic confinement. 
In the two panels there are the two frequencies 
$\Omega_{0,2}^{(a)}$ and $\Omega_{0,2}^{(b)}$ of the coupled monopole 
and quadrupole modes as a function of the number $N$ of atoms.
Filled circles with error bars: numerical results obtained solving 
Eqs. (\ref{continuo}) and (\ref{motion}) 
with Eq. (\ref{invert}) and $\lambda=0.25$.
Dashed lines: analytical results, i.e. 
Thomas-Fermi Eq. (\ref{tf-stringa2}). 
Universal parameter of the unitary Fermi gas: $\xi=0.42$.}
\label{fig2}
\end{figure}

For this reason we solve numerically the extended 
hydrodynamic equations (\ref{continuo}) and (\ref{motion}). 
In particular, by using our finite-difference predictor-corrector 
Crank-Nicolson code in real time \cite{sala-numerics}, we integrate a 
time-dependent nonlinear Schr\"odinger 
equation, which is fully equivalent (see \cite{flavio,flavio5}) 
to Eqs. (\ref{continuo}) and (\ref{motion}).

Fig. \ref{fig1} refers to the unitary Fermi gas under isotropic 
($\omega_{\rho}=\omega_z$) harmonic confinement. 
In the two panels we plot the monopole frequency 
$\Omega_{0}$ (upper panel) and the quadrupole frequency 
$\Omega_{2}$ (lower panel) as a function of the number $N$ of atoms. 
As expected \cite{castin}, the frequency $\Omega_0$ of the monopole 
mode does not depend on the number $N$ of particles and it is given by 
\beq 
\Omega_0 = 2 \omega_{\rho} \; . 
\label{exact-castin}
\eeq 
On the contrary, the figure shows that 
the frequency $\Omega_2$ of the quadrupole mode depends on $N$ 
and for large values of $N$ it approaches asymptotically 
the value $\sqrt{2}\omega_{\rho}$, 
appropriate to the case of neglecting
the gradient and backflow terms, see \cite{stringa}. 
Note that the filled circles are the results with $\lambda=0.25$ while the 
dashed lines show the analytical results \cite{castin,stringa}. 
Remarkably, for small values of $N$ the gradient term enhances the quadrupole 
frequency $\Omega_2$. In the isotropic case ($\omega_{\rho}=\omega_z$) 
the quadrupole frequency $\Omega_2$ in the limit $N\to \infty$ 
it gives the Thomas-Fermi result (i.e. without the gradiente term) 
\cite{stringa} 
\beq 
\Omega = \sqrt{2}\omega_{\rho} \; , 
\label{tf-stringa}
\eeq
while in the limit $N\to 0$ it gives 
$\Omega = 2 \omega_{\rho}$, which is the quadrupole oscillation frequency 
of non-interacting atoms (the same result holds for ideal 
fermions and ideal bosons) \cite{lipparini}. 

In Figs. \ref{fig2} we consider the unitary Fermi gas 
under anisotropic but axially-symmetric ($\omega_{\rho}=2\omega_z$) 
harmonic confinement. 
In this case monopole and quadrupole modes are coupled 
and we have determined numerically the two associated 
frequencies $\Omega_{0,2}^{(a)}$ 
and $\Omega_{0,2}^{(b)}$. Also in this case the gradient 
term increases the frequencies for small values of $N$. 
Moreover, for large values of $N$ these frequencies reduce to 
the results without gradient term \cite{stringa}  
\beq 
\Omega_{0,2}^{(a),(b)}= \sqrt{
\frac{5}{3}\omega_{\rho}^2 + {4\over 3}\omega_z^2 \pm {1 \over 3}
\sqrt{25\omega_{\rho}^4+16\omega_z^4-32\omega_{\rho}^2\omega_z^2} 
} \; , 
\label{tf-stringa2}
\eeq
which correspond to the dashed lines. 
Our calculations show that the frequency $\Omega_2$ of Fig. \ref{fig2}, 
and the frequencies $\Omega_{0,2}^{(a)}$ and $\Omega_{0,2}^{(b)}$ 
of Figs. \ref{fig2} give a clear signature 
of the presence of the von-Weizsacker gradient term. 

We stress that current experiments with ultracold atoms at unitarity 
can detect deviations from the Thomas-Fermi approximation,  
as done some years ago for Bose-Einstein 
condensates \cite{jin}. 

\section{Inclusion of a backflow term}

Inspired by the papers of Son and Wingate \cite{son} 
and Manes and Valle \cite{valle} in this section we consider the 
inclusion of a backflow term in the extended superfluid Lagrangian. 
This backflow term depends on the velocity strain, as suggested for superfluid 
$^4$He many years ago by Thouless \cite{less} 
and more recently by Dalfovo and collaborators \cite{treiner}. 
In particular, we consider the Lagrangian density 
\beq 
\mathscr{L} = \mathscr{L}_0 + \mathscr{L}_W + \mathscr{L}_B \; , 
\label{lagrangian1}
\eeq
where $\mathscr{L}_0$ and $\mathscr{L}_W$ are given 
by Eqs. (\ref{l0}) and (\ref{lw}) respectively, 
and the backflow term $\mathscr{L}_B$ reads 
\beq 
\mathscr{L}_B = - {\hbar^2\over m} n^{1/3} 
[\gamma_1 (\nabla^2\theta)^2+\gamma_2 (\partial_i\partial_j\theta)^2] \; . 
\eeq
Notice that $i,j=x,y,z$ and  summations over repeated indices are implied. 
Again, for a generic superfluid the parameters 
$\gamma_1$ and $\gamma_2$ of the backflow term must be determined from 
microscopic calculations or from comparison with experimental data. 

The Lagrangian density (\ref{lagrangian1}) depends on the dynamical variables 
$\theta({\bf r},t)$ and $n({\bf r},t)$. 
The conjugate momenta of these dynamical variables are then given by 
\beqa 
\pi_{\theta} &=& {\partial \mathscr{L} \over \partial {\dot \theta}} 
= - \hbar \, n \; , 
\\
\pi_{n} &=& {\partial \mathscr{L} \over \partial {\dot n}} = 0 \; , 
\eeqa
and the corresponding Hamiltonian density reads 
\beq 
\mathscr{H} = \pi_{\theta} \,  {\dot \theta} + \pi_{n} \,  {\dot n} 
- \mathscr{L} =  - \hbar \, n \,  {\dot \theta} - \mathscr{L} \; , 
\eeq
namely 
\beqa 
\label{hamiltonian}
\mathscr{H} &=& {\hbar^2\over 2m} ({\boldsymbol \nabla} \theta)^2 \, n 
 + U({\bf r}) \, n + {\cal E}_0(n) 
\\
&+& \lambda {\hbar^2\over 8m} {({\boldsymbol\nabla} n)^2\over n} + 
{\hbar^2\over m} n^{1/3} 
[\gamma_1(\nabla^2\theta)^2+\gamma_2(\partial_i\partial_j\theta)^2] \; ,  
\nonumber
\eeqa
which is the sum of the flow kinetic energy density 
$\hbar^2({\boldsymbol \nabla} \theta)^2n/(2m)
= (1/2)m v^2n$, the external energy density $U({\bf r})n$, 
the internal energy density 
${\cal E}_0(n)$ without the gradient correction, 
the gradient correction 
$\lambda ({\hbar^2/ 8m}) {({\boldsymbol\nabla} n)^2/n}$ 
to the internal energy, and the backflow energy density 
$({\hbar^2/m}) n^{1/3} 
[\gamma_1(\nabla^2\theta)^2+\gamma_2(\partial_i\partial_j\theta)^2]
=mn^{1/3}[\gamma_1({\boldsymbol\nabla} \cdot {\bf v})^2
+\gamma_2(\partial_iv_j)^2]$. 

The Hamiltonian density (\ref{hamiltonian}) is nothing else than 
the energy density recently found by Manes and Valle \cite{valle} 
with a derivative expansion from their effective field theory 
of the the Goldstone field \cite{son,valle}. 
The effective field theory of Manes and Valle \cite{valle} 
traces back to the old hydrodynamic results of Popov \cite{popov} 
and generalizes the one 
derived by Son and Wingate \cite{son} for the unitary Fermi gas 
from general coordinate invariance and conformal invariance. 
Actually, at next-to-leading order 
Son and Wingate \cite{son} found an additional term proportional 
to $\nabla^2U({\bf r})$, which has been questioned by Manes and Valle 
\cite{valle} and which is absent in our approach. In addition, 
Manes and Valle \cite{valle} have stressed that the conformal 
invariance displayed by the unitary Fermi gas implies 
\beq 
\gamma_2=-3\gamma_1 \; .  
\label{ziobilly}
\eeq 
Note that a paper of Schakel \cite{schakel} 
confirms the results of Manes and Valle. 

We are interested on the propagation of sound waves in superfluids. 
For simplicity we set $U({\bf r})= 0$, 
and consider a small fluctuation $\phi({\bf r},t)$ 
of the phase $\theta({\bf r},t)$ around the stationary phase 
$\theta_{0}(t)= -({\mu/\hbar}) t$, namely 
\beq 
\phi({\bf r},t) = \theta({\bf r},t) - \theta_0(t) \; , 
\eeq
and a small fluctuation ${\rho}({\bf r},t)$ of the density 
$n({\bf r},t)$ around the constant and uniform density $n_0$, namely 
\beq
{\rho}({\bf r},t) = n({\bf r},t) - n_0  \; . 
\eeq
From the full Lagrangian density (\ref{lagrangian1}) it is then quite 
easy to find the quadratic Lagrangian density $\mathscr{L}^{(2)}$ 
of the fluctuating fields 
$\phi({\bf r},t)$ and $\rho({\bf r},t)$:
\beqa 
\label{lagrangian2}
\mathscr{L}^{(2)} &=& - \hbar \, {\dot \phi} \, \rho - 
{\hbar^2n_0\over 2m} ({\boldsymbol \nabla} \phi)^2   
- {mc_s^2\over 2n_0} \rho^2 
\\
&-& 
\lambda {\hbar^2\over 8m\, n_0} ({\boldsymbol\nabla} \rho)^2
- \gamma {\hbar^2n_0^{1/3}\over m} (\nabla^2\phi)^2 \; ,  
\nonumber
\eeqa
where $c_s$ is the sound velocity of the generic superfluid, given by 
\beq 
c_s^2 = 
{n_0\over m} {\partial^2 {\cal E}_0(n_0)\over \partial n^2} \; ,   
\label{def-cs}
\eeq
and $\gamma=\gamma_1+\gamma_2$. In fact, $(\nabla^2\theta)^2$ and 
$(\partial_i\partial_j\theta)^2$ 
differ by a total derivative \cite{valle} and 
consequently, since at the quadratic order 
the coefficients in front of them are constants, 
one derives Eq. (\ref{lagrangian2}) with $\gamma = \gamma_1+\gamma_2$. 
The linear equations of motion associated to 
the quadratic Lagrangian $\mathscr{L}_2$ read 
\beqa
{\partial \over \partial t}
\rho + n_0 {\boldsymbol\nabla} \cdot {\bf v} 
- {2} n_0^{1/3} \gamma \, 
\nabla^2 ({\boldsymbol\nabla}\cdot {\bf v}) = 0 \; , 
\\
{\partial\over \partial t} {\bf v} + {c_s^2\over n_0} 
{\boldsymbol \nabla} \rho - {\lambda \hbar^2\over 4m^2n_0} 
{\boldsymbol \nabla}( \nabla^2\rho ) = 0 \; ,  
\eeqa
with ${\bf v}=(\hbar/m)\nabla\phi$. 
These equations can be arranged in the form of 
the following wave equation
\beqa 
\label{dispersion}
\Big[ {\partial^2\over\partial t^2} - c_s^2 \nabla^2 
&+& \Big( \lambda {\hbar^2\over 4m^2}  + \gamma
{2c_s^2\over n_0^{2/3}}\Big) \nabla^4 
\\
&-& \lambda \gamma {\hbar^2\over 2m^2n_0^{2/3}} 
\nabla^6 \Big] \rho({\bf r},t) = 0 \; . 
\nonumber
\eeqa
This wave equation admits  
monochromatic plane-wave solutions, 
where the frequency $\omega$ and the wave vector ${\bf q}$ are 
related by the dispersion formula $\omega=\omega(q)$ given by 
\beq 
\label{w-dis}
\hbar \, \omega(q) = \sqrt{\Big({\hbar^2q^2\over 2m}+\gamma 
{\hbar^2q^4\over mn_0^{2/3}}\Big)
\Big(\lambda{\hbar^2q^2\over 2m}+2mc_s^2\Big)} \; . 
\eeq
Notice that a negative value of $\gamma$ implies that 
the frequency $\omega(q)$ becomes imaginary 
for $q > n_0^{1/3}/\sqrt{2|\gamma|}$. However, $\gamma$ is expected 
to be very small and the hydrodynamics is no loger valid for these
large values of $q$. 

It is instead useful to expand $\omega(q)$ for small values of $q$ 
(long-wavelength hydrodynamic regime), finding 
\beq 
\hbar \, \omega(q) 
= c_s \, \hbar q + {\hbar \over 2} \big(  
\lambda {\hbar^2\over 4m^2c_s}
+\gamma {2c_s\over n_0^{2/3}} \big) \, q^3  + ... \;  . 
\eeq
The dispersion relation is linear in $q$ only for small values of the 
wavenumber $q$ and the coefficient of cubic correction depends 
on a combination of the gradient parameter $\lambda$ 
and backflow parameter $\gamma$. For $\gamma =0$ one recovers 
the dispersion relation 
we have proposed some years ago \cite{flavio}, while setting also 
$\lambda=0$ one gets the familiar linear dispersion relation 
$\omega=c_s \, q$ of phonons. 
In the case of the unitary Fermi gas one has 
\beq 
c_s^2 = {\hbar^2\over m^2} {\xi \over 3} (3\pi^2)^{2/3} n_0^{2/3} \; .  
\label{sound-unit}
\eeq
Moreover, we have seen that the backflow parameters 
are related by the formula (\ref{ziobilly}), which means 
\beq 
\gamma=\gamma_1+\gamma_2=-2\gamma_1 \; . 
\eeq
Consequently, at the cubic order in $q$ Eq. (\ref{w-dis}) gives 
\beq 
{\omega(q) \over c_s k_F} = {q\over k_F}  + \Gamma \, {q^3\over k_F^3} \; , 
\label{disprosio}
\eeq
where $k_F=(3\pi^2n_0)^{2/3}$ is the Fermi wavenumber and 
\beq 
\Gamma = {3\lambda \over 8\xi} - 2 (3\pi^2)^{2/3} \gamma_1 \; .  
\eeq
Within a mean-field approximation Manes and Valle \cite{valle} 
have found $\gamma_1 \simeq 0.006$, which implies $\gamma \simeq -0.012$ 
and $\Gamma \simeq 0.12$, using $\xi=0.4$ and $\lambda=0.25$. 
As recently discussed by Mannarelli, Manuel and Tolos \cite{tolos}, 
the sign of $\Gamma$ has a dramatic effect on the possible phonon 
interaction channels: the three-phonon Beliaev process, i.e. 
the decay of a phonon into two phonons \cite{beliaev}, 
is allowed only for positive values of $\Gamma$. Under this condition 
($\Gamma \geq 0$) the phonon has a finite 
life-time and the frequency $\omega(q)$ 
possesses an imaginary part $\mbox{Im}[{\omega(q)}]$
due to this three-phonon decay \cite{beliaev,lev}. In particular, we find  
\beq 
\mbox{Im}[{\omega(q)}] = - {\hbar q^5\over 270 \, \pi \, m \, n_0} \; . 
\label{beliaev}
\eeq
This formula of Beliaev damping is easily derived 
from Beliaev theory \cite{beliaev,lev} taking into 
account Eq. (\ref{sound-unit}). 

It is important to point out that 
the sign of $\Gamma$ in Eq. (\ref{disprosio}) 
was debated also without the backflow term. 
In 1998 Marini, Pistolesi and Strinati \cite{marini} 
found $\Gamma>0$ at unitarity 
by including Gaussian fluctuations to the mean-field BCS-BEC crossover. 
In 2005 Combescot, Kagan and 
Stringari \cite{combescot} derived Eq. (\ref{disprosio}) 
with a negative $\Gamma$ at unitarity on the basis of a dynamical BCS model. 
In 2011 Schakel \cite{schakel} obtained a positive $\Gamma$ 
at unitarity by using a derivative expansion technique, finding exactly 
the values of $\Gamma$ predicted by Ref. \cite{marini} 
in the full BCS-BEC crossover.

To conclude this section, we observe that, for a generic many-body system, 
the dispersion relation can be written as \cite{lipparini}
\beq 
\hbar \, \omega(q) = \sqrt{m_1(q)\over m_{-1}(q)} 
\; , 
\eeq
where $m_n(q)$ is the $n$ moment of the dynamic structure function 
$S(q,\omega)$ of the many-body system under investigation, 
namely \cite{lipparini}
\beq 
m_n(q) = \int_0^{\infty} d\omega \ S(q,\omega) \ (\hbar \omega)^n \; . 
\eeq 
In our problem, Eq. (\ref{dispersion}), it is straightforward 
to recognize (see also \cite{treiner}) that 
\beq 
m_1(q) = {\hbar^2q^2\over 2m} +\gamma {\hbar^2 q^4 \over mn_0^{2/3}}
\eeq
and 
\beq 
m_{-1}(q) = {1 \over  \lambda {\hbar^2q^2\over 2m} + 2 m c_s^2} \; .  
\eeq
In general, the static response function $\chi(q)$ is defined as 
\cite{lipparini}
\beq 
\chi(q) = -2 \ m_{-1}(q) \; ; 
\eeq 
in our problem it reads: 
\beq 
\chi(q) = - {2 \over  \lambda {\hbar^2q^2\over 2m} + 2 m c_s^2} \; ,    
\eeq
which satisfies the exact sum rule $\chi(0)=-1/mc_s^2$ \cite{lipparini}. 
The static structure factor $S(q)$, defined as \cite{lipparini}
\beq 
S(q) = m_0(q) = \int_0^{\infty} d\omega \ S(q,\omega) \; ,   
\eeq
can be approximated by the expression 
\beq 
S(q) = \sqrt{m_1(q)\, m_{-1}(q)} = 
\sqrt{{\hbar^2q^2\over 2m}  +\gamma {\hbar^2q^4
\over mn_0^{2/3}} \over  \lambda {\hbar^2q^2\over 2m} + 2 m c_s^2}
\; , 
\eeq
which gives an upper bound of $S(q)$ \cite{lipparini} 
and reduces to $S(q)=\hbar q/(2mc_s)$ for small $q$. 

Finally, we remark that one can also calculate the frequencies $\Omega$ 
of collective oscillations of the unitary Fermi gas 
under the action of the trapping potential given by 
Eq. (\ref{oscillator}) taking into account the backflow. 
We have verified that in the case of spherically-symmetric harmonic 
confinement ($\omega_{\rho}=\omega_{z}$) the monopole mode $\Omega_0$ 
is not affected by the backflow term, i.e. $\Omega_0=2\omega_{\rho}$. 
Moreover, for large values of $N$ the contribution due to the 
backflow becomes negligible, similarly to the von Weizs\"acker one. 

\section{Conclusions}

We have calculated collective modes of the anisotropic unitary Fermi gas 
by using the equations of extended superfluid hydrodynamics. 
In particular, we have shown that a gradient correction of the von-Weizsacker 
form in the hydrodynamic equations strongly affects the frequencies 
of collective modes of the system under axially-symmetric anisotropic 
harmonic confinement. We have found that, for both monopole and quadrupole 
modes, this effect becomes negligible only in the regime 
of a large number of fermions, where one recovers the 
predictions of superfluid hydrodynamics \cite{stringa}. 
In the last part of the paper 
we have considered the inclusion of a backflow term 
in the extended hydrodynamics of superfluids. 

We believe our results can trigger the interest of experimentalists. 
Some years ago beyond-Thomas-Fermi effects due to the dispersive 
gradient term have been observed by measuring the frequencies of 
collective modes in trapped Bose-Einstein condensates \cite{jin}. 
Moreover, the spectrum of phonon excitations and Beliaev decay 
have been observed in a quasi-uniform Bose-Einstein condensate 
with Bragg pulses \cite{nir}. 
Performing similar measurements in the unitary Fermi gas 
can shed light on the role played by gradient and backflow corrections 
in the superfluid hydrodynamics. 

\section*{Acknowledgments}

LS and FT thank University of Padova (Research Project 
"Quantum Information with Ultracold Atoms in Optical Lattices"), 
Cariparo Foundation (Excellence Project 
"Macroscopic Quantum Properties of Ultracold Atoms under 
Optical Confinement"), and MIUR (PRIN Project "Collective Quantum Phenomena: 
from Strongly-Correlated Systems to Quantum Simulators") 
for partial support.


\begin{thebibliography}{99.}

\bibitem{flavio} L. Salasnich and F. Toigo, 
Phys. Rev. A {\bf 78}, 053626 (2008). 

\bibitem{flavio2} L. Salasnich, Laser Phys. {\bf 19}, 642 (2009). 

\bibitem{flavio3} F. Ancilotto, L. Salasnich, and F. Toigo, 
Phys. Rev. A {\bf 79}, 033627 (2009). 

\bibitem{flavio4} S.K. Adhikari and L. Salasnich, 
New J. Phys. {\bf 11}, 023011 (2009). 

\bibitem{flavio5} L. Salasnich, F. Ancilotto, and F. Toigo, 
Laser Phys. Lett. {\bf 7}, 78 (2010). 

\bibitem{flavio6} L. Salasnich, EPL {\bf 96}, 40007 (2011). 

\bibitem{flavio7} F. Ancilotto, L. Salasnich, and F. Toigo, 
Phys. Rev. A {\bf 85}, 063612 (2012). 

\bibitem{flavio8} L. Salasnich, 
Few-Body Syst. {\bf 54}, 697 (2013).

\bibitem{von} C.F. von Weizs\"acker, Zeit. Phys. {\bf 96}, 431 (1935).

\bibitem{tosi} N.H. March and M. P. Tosi, 
Proc. R. Soc. A {\bf 330}, 373 (1972). 

\bibitem{tso} E. Zaremba and H.C. Tso, 
Phys. Rev. B {\bf 49}, 8147 (1994).

\bibitem{nick} N. Manini and L. Salasnich, 
Phys. Rev. A, {\bf 71}, 033625 (2005); 
G. Diana, N. Manini, and L. Salasnich, 
Phys. Rev. A, {\bf 73}, 065601 (2006). 

\bibitem{kim} Y.E. Kim and A.L. Zubarev, 
Phys. Rev. A {\bf 70}, 033612 (2004). 

\bibitem{v2} M.A. Escobedo, M. Mannarelli and C. Manuel, 
Phys. Rev. A {\bf 79}, 063623 (2009). 

\bibitem{v3} E. Lundh and A. Cetoli, 
Phys. Rev. A  {\bf 80}, 023610 (2009). 

\bibitem{v4} G. Rupak and T. Sch\"afer, 
Nucl. Phys. A {\bf 816}, 52 (2009). 

\bibitem{v5} S.K. Adhikari, Laser Phys. Lett. {\bf 6}, 901 (2009). 

\bibitem{v6} W.Y. Zhang, L. Zhou, and Y.L. Ma, 
EPL {\bf 88}, 40001 (2009). 

\bibitem{v7} A. Csordas, O. Almasy, and P. Szepfalusy, 
Phys. Rev. A {\bf 82}, 063609 (2010). 

\bibitem{v8} S. N. Klimin, J. Tempere, and J.P.A. Devreese, 
J. Low Temp. Phys. {\bf 165}, 261 (2011). 

\bibitem{less} D.J. Thouless, Ann. Phys. {\bf 52}, 403 (1969). 

\bibitem{treiner} F. Dalfovo, A. Lastri, L. Pricaupenko, 
S. Stringari, and J. Treiner, Phys. Rev. B {\bf 52}, 1193 (1995). 

\bibitem{beliaev} S.T. Beliaev, Sov. Phys. JETP {\bf 7}, 299 (1958). 

\bibitem{popov} V.N. Popov, {\it Functional Integrals in 
Quantum Field Theory and Statistical Physics} 
(Reidel, Dordrecht, 1983). 

\bibitem{sala-numerics} E. Cerboneschi, R. Mannella, E. Arimondo, 
and L. Salasnich, Phys. Lett. A {\bf 249}, 495 (1998); 
G. Mazzarella and L. Salasnich,  
Phys. Lett. A {\bf 373}, 4434 (2009). 

\bibitem{castin} Y. Castin, Comptes Rendus Physique {\bf 5}, 407 (2004).

\bibitem{forzati} Y.-H. Hou, L.P. Pitaevskii, and S. Stringari, 
Phys. Rev. A. {\bf 87}, 033620 (2013). 

\bibitem{stringa} M Cozzini, S. Stringari, Phys. Rev. Lett. {\bf 91}, 
070401 (2003). 

\bibitem{jin} D.S. Jin, J. R. Ensher, M. R. Matthews, C. E. Wieman, and E.
A. Cornell, Phys. Rev. Lett. {\bf 77}, 420 (1996). 

\bibitem{son} D.T. Son and M. Wingate, Ann. Phys. (N.Y.) {\bf 321}, 197 (2006). 

\bibitem{valle} J.L. Manes and M.A. Valle, 
Ann. Phys. (N.Y.) {\bf 324}, 1136 (2009). 

\bibitem{schakel} A.M.J. Schakel, 
Ann. Phys. (N.Y.) {\bf 326}, 193 (2011). 

\bibitem{tolos} M. Mannarelli, C. Manuel, and L. Tolos, 
e-preprint arXiv:12.5152. 

\bibitem{lev} L.P. Pitaevskii and S. Stringari, 
Bose-Einstein Condensation (Oxford Univ. Press, 
Oxford, 2003), pp. 72-74.   

\bibitem{lipparini} E. Lipparini, {\it Modern Many-Particle Physics: 
Atomic Gases, Nanostructures and Quantum Liquids} 
(World Scientific, Singapore, 2008). 

\bibitem{marini} M. Marini, F. Pistolesi, and G.C. Strinati, 
Eur. Phys. J B {\bf 1}, 151 (1998). 

\bibitem{combescot} R. Combescot, M.Yu. Kagan, and S. Stringari, 
Phys. Rev. A {\bf 74}, 042717 (2006). 

\bibitem{nir} E.E. Rowen, N. Bar-Gill, and N. Davidson, 
Phys. Rev. Lett. {\bf 101}, 010404 (2008); 
E.E. Rowen, N. Bar-Gill, R. Pugatch, and N. Davidson, 
Phys. Rev, A {\bf 77}, 033602 (2008). 

\end{thebibliography}
\end{document}